\documentclass[article,nojss]{jss}

\usepackage{amsmath, amssymb, amsthm}
\usepackage{graphicx}

\usepackage{thumbpdf,lmodern}

\usepackage{framed}

\newcommand{\class}[1]{`\code{#1}'}

\let\oldcode\code
\renewcommand{\code}[1]{\textrm{\oldcode{#1}}}

\newcommand{\nay}{\code{FALSE}}
\newcommand{\yay}{\code{TRUE}}
\newcommand{\NA}{\code{NA}}

\usepackage{todonotes}

\author{Mark P.J. van der Loo\\Statistics Netherlands
   \And Edwin de Jonge\\Statistics Netherlands}
\Plainauthor{Mark P.J. van der Loo, Edwin de Jonge}

\title{Data Validation Infrastructure for \proglang{R}}
\Plaintitle{Data validation infrastructure for R}
\Shorttitle{Data validation infrastructure for R}

\Abstract{
Checking data quality against domain knowledge is a common activity that
pervades statistical analysis from raw data to output.  The \proglang{R}
package \pkg{validate} facilitates this task by capturing and applying expert
knowledge in the form of validation rules: logical restrictions on variables,
records, or data sets that should be satisfied before they are considered valid
input for further analysis. In the \pkg{validate} package, validation rules are
objects of computation that can be manipulated, investigated, and confronted
with data or versions of a data set. The results of a confrontation are then
available for further investigation, summarization or visualization. Validation
rules can also be endowed with metadata and documentation and they may be
stored or retrieved from external sources such as text files or tabular
formats. This data validation infrastructure thus allows for systematic,
user-defined definition of data quality requirements that can be reused for
various versions of a data set or by data correction algorithms that are
parameterized by validation rules.}

\Plainkeywords{data checking, data quality, data cleaning, R}

\Address{
  Mark van der Loo, Edwin de Jonge\\
  Research and Development\\
  Statistics Netherlands\\
  Henri Faasdreef 312 \\
  2492JP Den Haag, The Netherlands\\
  E-mail: \email{m.vanderloo@cbs.nl}, \email{e.dejonge@cbs.nl}\\
}

\usepackage{Sweave}
\begin{document}

\textbf{Citation.} MPJ van der Loo and E de Jonge (2019), \emph{Data Validation Infrastructure for R},
Journal of Statistical Software (accepted for publication).

\section{Introduction}
\label{sect:introduction}
Checking whether data satisfy assumptions based on domain knowledge pervades
data analyses. Whether it is raw data, cleaned up data, or output of a
statistical calculation, data analysts have to scrutinize data sets at every
stage to ensure that they can be used for reporting or further computation.  We
refer to this procedure of investigating the quality of a data set and deciding
whether it is fit for purpose as a `data validation' procedure.

Many things can go wrong while creating, gathering, or processing data.
Accordingly there are many types of checks that can be performed. One usually
distinguishes between technical checks that are related to data structure and
data type, and checks that are related to the topics described by the data.
Examples of technical checks include testing whether an `age' variable is
numeric, whether all necessary variables are present, or whether the
identifying variable (primary key) is unique across records.  Examples of tests
that relate to subject matter knowledge include range checks (the age of a
person should be between 0 and say, 120), and consistency checks between
variables (a person under the age of 18 can not be married). Some checks
require simple comparisons, while others involve elaborate calculations.  For
example consider a check where we test that the mean profit to turnover ratio
in a certain economic sector has not changed more than 10\% between the current
quarter and the same quarter last year. To execute such a check we need data on
profit and turnover for two quarters, compute their ratios (taking exceptions
such as dividing by zero into account), average them, and then compare the
relative absolute difference with a fixed value (0.1).

Since data validation is such a common and diverse task, it is desirable to use
a dedicated tool that allows one to systematize and organize data validation
procedures. There are several recent contributions to the \proglang{R} language
\citep{rcore} that aim to support this in a systematic way.  The
\pkg{pointblank} package of \cite{iannone2018pointblank} allows users to test
data against a number of hard-coded validation checks. For example, there are
separate functions to determine whether values in a column are positive,
nonnegative, or in a certain range.  The \pkg{dataMaid} package
\citep{petersen2018datamaid} creates generic summary reports and figures that
are aimed at detecting common problems with data.  Detected problems include
the presence of special values (such as \code{Inf}), empty strings and
identifying potential outliers.  \citet{fischetti2018assertr} published the
\pkg{assertr} package, which allows for checking data against a number of
predefined (and parameterized) data checks called `predicates'.  It is
specifically geared for use with a function composition operator such as the
\pkg{magrittr} `forward pipe' operator of \citet{bache2014magrittr}. Depending
on the outcome of a predicate a custom function can be evaluated and the
utility of \pkg{assertr} is therefore comparable to a try-catch mechanism.  The
\pkg{editrules} package \citep{jonge2015editrules} allows users to specify
their own checks (called `edit rules') and apply them to data.  The checks that
can be specified are limited to simple in-record linear equalities and
inequalities over the variables, and certain conditional checks (e.g.,
\texttt{IF age < 18 THEN married == FALSE}).

The \pkg{validate} package \citep{loo2018validate} presented in this paper
takes an approach that is more general than the packages mentioned above.
Specifically, \pkg{validate} does not limit the type of checks by hard-coding
them or by limiting them to certain rule types. Instead it provides a
convenient domain specific language (DSL) in which the user can express any
type of validation task in the form of any number of `validation rules'. The
package takes care of reading, parsing and applying the rules to the data and
offers tools to analyze and visualize the results.  Moreover, it treats these
validation rules as first-class citizens so they can be organized, manipulated,
and analyzed like any other data type. 

Defining a language specifically for the definition of validation rules has
several advantages. First, a formal language necessarily implements a strict
demarcation of the concept of `data validation'. A data validation language
therefore incites users to separate their thinking about data quality
requirements from the data processing work flow.  Second, expressing data
quality requirements in the form of a formal data validation language allows
for unambiguous and clear communication of data quality or data quality
requirements between data producers and consumers.  Third, it allows for
systematic development and maintainance of data quality requirements based on
established techniques for developing source code. This includes practices like
testing, reviewing, documenting, and version control. Fourth, it allows for the
reuse of validation rules for purposes other than quality measurement.
Important examples include the use of algorithms that can automatically find
errors and adjust data to satisfy the validation rules \citep{waal2011handbook,
loo2018statistical}. Some examples of this are reported in
Section~\ref{sect:demo}. Finally, treating validation rules as enities on their
own opens up interesting new avenues of inquiry including questions like:
`which rules were violated most often?' and `what variables were involved in
these rules?' Such statistics can provide valuable clues to issues related to
data gathering and processing that occurred prior to data validation.  

The current version of the package is limited to validating data that is
represented as records in an \proglang{R} data frame. This means that in its
current form other useful data types including networks, geographical data,
time series or (high-dimensional) arrays are excluded. However, the package is
extendable so it is possible to add such functionality without altering its
associated workflow or syntax.

In the remainder of this paper we first give a more precise definition of data
validation (Section~\ref{sect:datavalidation}).  Next, in
Section~\ref{sect:validate} we discuss how this definition is implemented in a
domain specific language and demonstrate the basic data validation work flow
with the package.  We also demonstrate how \pkg{validate} treats validation as
first-class citizens, allowing users to create, read, update, and delete (CRUD)
rules interactively.  Rules can also be endowed with metadata such as names and
descriptions and we discuss how rules can be imported from, and exported to
several formats.  Section~\ref{sect:validate} also discusses options such has
how to control the accuracy when testing numerical equalities during data
validation, and how to analyse and visualise the results of a data validation.
In Section~\ref{sect:implementation} we provide some background and discuss the
object model that is implemented by the package. In Section~\ref{sect:demo} we
demonstrate how \pkg{validate} can be used to both control and monitor a small
data cleaning task. We conclude with a summary and outlook in
Section~\ref{sect:conclusion}.

\section{Data validation}
\label{sect:datavalidation}
Intuitively, data validation is a decision-making process where, based on data
investigations, one decides whether the data set is fit for its intended
purpose. This notion is made more precise in the following definition, which is
currently the operational definition for the European Statistical
System\footnote{The European Statistical System (ESS) is a partership between
Eurostat (the European Union statistical authority) and statistical authorities
of member states of the European Union, the European Free Trade Association
(EFTA) and the European Economic Area (EEA). See
\href{https://ec.europa.eu/eurostat/web/european-statistical-system}{https://ec.europa.eu/eurostat/web/european-statistical-system}} \citep{zio2015methodology}.
\begin{quote}
Data validation is an activity in which it is verified whether or not a
combination of values is a member of a set of acceptable value combinations.
\end{quote}
In other words, one considers the collection of all datasets that might be
observed, and defines data validation as a procedure that selects the datasets
that are acceptable for further use. This definition is general enough to
permit a precise formal definition but also includes the option to approve data
by expert review. 

To develop the formal side, the following definition is sufficient for our
purpose \citep{loo2019data}.
\begin{quote}
A \emph{data validation function} is a function $v$ that accepts a data set and
returns an element in $\{\yay{}, \nay{}\}$. There is at least one
data set $s$ for which $v(s)=\yay{}$ and at least one data set $s'$ for which
$v(s')=\nay{}$.
\end{quote}
We say that $s$ \emph{passes} $v$ if $v(s)=\yay{}$ and that $s$ \emph{fails}
$v$ if $v(s)=\nay{}$. The definition requires that validation functions are
surjective on $\{\yay{},\nay{}\}$ for the following reasons. On one hand, if
$v(s)=\yay{}$ for every possible data set $s$, it does not distinguish between
valid and invalid data (a clear property demanded by the ESS definition). On
the other hand, if $v(s)=\nay{}$ for each possible data set $s$, then $v$ must
be a contradiction since no data can ever satisfy the demand expressed by $v$.

It is possible to define the concept of `each possible data set' formally by
defining data sets as collections of key-value pairs in a precise way.  For a
complete formal treatment of validation functions and some of their general
properties we refer the reader to \citet{loo2019data},
\citet[Section~5]{zio2015methodology} or \citet[Chapter~6]{loo2018statistical}.
For now, it is important to note that the definition does not pose any
restriction on the form of data set that is under scrutiny. In the current
definition a data set is viewed as an unstructured set of key-value pairs and
we do not impose a topology such as relational or network structure.

\subsection{Validation functions and validation rules}
A validation function can be defined by fixing a set of restrictions on a data
set. For example consider a data set with variables \code{age} and
\code{employment}. Before using the data set in a statistical analysis we wish
to ensure that age is in the range $[0,120]$ and that persons under the age of
15 do not have a paid job (have no employment). Knowing to what population the
data pertains, we also find it highly implausible to have an unemployment
fraction exceeding 30\%.  These demands can be expressed as a rule set
\begin{equation*}
\left\{
\begin{array}{l}
  \code{age} \geq 0 \textrm{ for all ages}\\
  \code{age} \leq 120 \textrm{ for all ages}\\
  \code{IF employment} == \code{"employed" THEN age}\geq 15 \textrm{ for all age-employment combinations}\\
  (\textrm{nr. of records with } \code{employment}== \code{"unemployed"})/(\textrm{nr. of records})\leq 0.3,
\end{array}\right.
\end{equation*}
The corresponding validation function evaluates each rule on a data set and
returns the logical conjunction (\code{AND}) of their values. 

Each restriction is ultimately a logical predicate, with propositions that
possibly consist of complex calculations on the data. In \pkg{validate} this is
implemented by allowing users to define a set of restrictions in the form of
\proglang{R} expressions that evaluate to a scalar or vector of type
\code{logical}. The package then takes care of confronting rules with data and
of the administration of the rules and results.

\section{Data validation with validate}
\label{sect:validate}
\begin{figure}
\includegraphics[width=\textwidth]{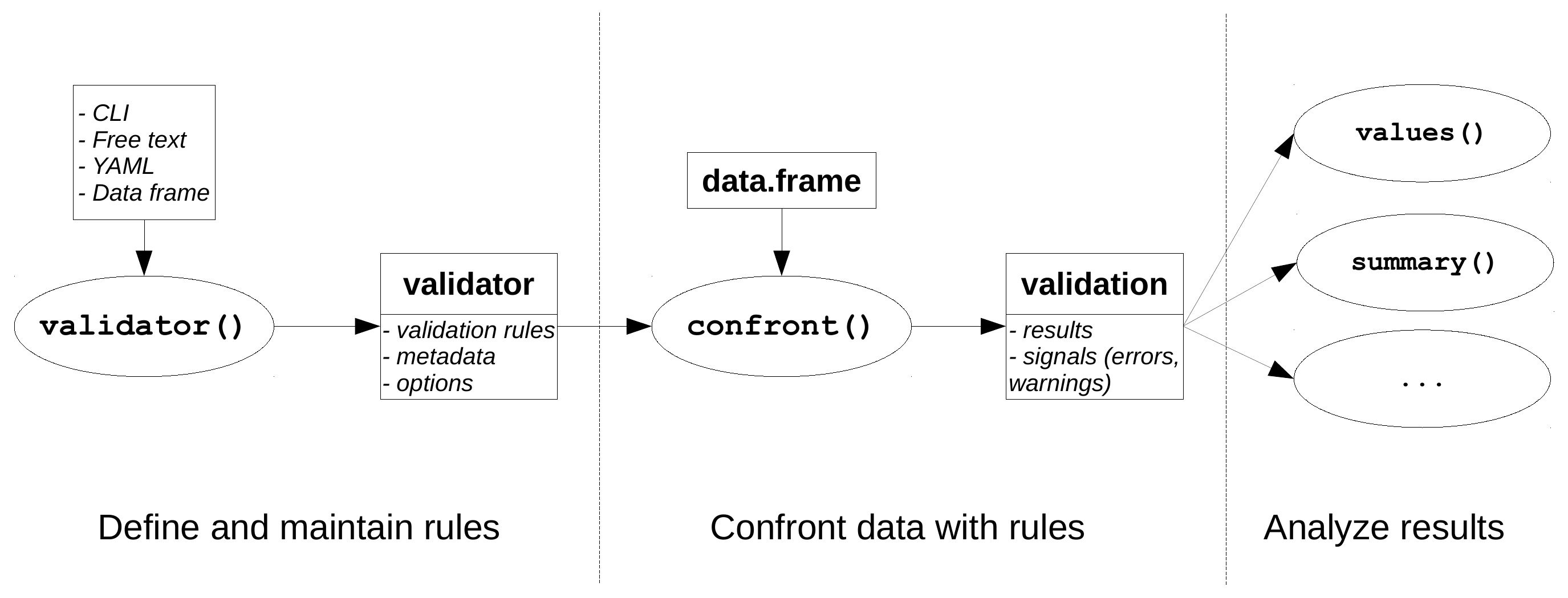}
\caption{Overview of the main workflow and objects in \pkg{validate}.}
\label{fig:workflow}
\end{figure}

A data validation task can be split up in  three consecutive subtasks: loading
the data and the validation rules, confronting the data with the rules, and
transforming and analyzing the results.  In what follows, we first demonstrate
this workflow with an example.  Next, detailed descriptions of the most
important features of the package are given in
Sections~\ref{sect:dsl}--\ref{sect:confrontation}. An overview of the
workflow is also given in Figure~\ref{fig:workflow}.

In the following we use the \code{retailers} data set included with the
package. This data set consists of 60 records with information on supermarkets,
such as the number of staff employed, several income amounts and several cost
items.
\begin{Schunk}
\begin{Sinput}
R> library("validate")
R> data("retailers")
R> head(retailers[3:8], 3)
\end{Sinput}
\begin{Soutput}
  staff turnover other.rev total.rev staff.costs total.costs
1    75       NA        NA      1130          NA       18915
2     9     1607        NA      1607         131        1544
3    NA     6886       -33      6919         324        6493
\end{Soutput}
\begin{Sinput}
R> retailers$id <- sprintf("RET%2d",1:nrow(retailers))
\end{Sinput}
\end{Schunk}
In the last expression we add a primary key to be better able to identify
results later.  Having access to a unique record identifier is not strictly
necessary but it does make connecting validation results to original data
easier later on in this example.

To test this data we first capture a set of restrictions in a \class{validator}
object, using a function of the same name. We start with a few rules defined
on the command-line.
\begin{Schunk}
\begin{Sinput}
R> rules <- validator(
+     st    = staff     >= 0
+   , to    = turnover  >= 0
+   , or    = other.rev >= 0
+   , st.cs = if (staff > 0 ) staff.costs > 0
+   , bl    = turnover + other.rev == total.rev
+   , mn    = mean(profit, na.rm=TRUE) >= 1
+ )
\end{Sinput}
\end{Schunk}
The first three rules are non-negativity restrictions.  The fourth rule
indicates that when a retailer has employees (staff), there must be associated
costs. The fifth rule is an account balance check and the last rule expresses
that perhaps some individual retailers have negative profit (loss), but the
sector as a whole is expected to be profitable. Each rule is given a (very)
short name. For instance the first rule is named \code{"st"}.

To test the data against these rules, we use the function \code{confront}.
\begin{Schunk}
\begin{Sinput}
R> confront(retailers, rules, key="id")
\end{Sinput}
\begin{Soutput}
Object of class 'validation'
Call:
    confront(dat = retailers, x = rules, key = "id")

Confrontations: 6
With fails    : 2
Warnings      : 0
Errors        : 0
\end{Soutput}
\end{Schunk}
The resulting object holds validation results and some metadata on the data
validation procedure. When printed, it shows the number of `confrontations'
performed (there are six rules) and number of confrontations that resulted in
at least one failure. The number of warnings and errors do not refer to data
failing a rule. Rather, they show whether trying to execute a rule raised a
warning or error. An error occurs when a rule cannot be executed, for example
when it refers to a variable that is not in the data set.
\begin{Schunk}
\begin{Sinput}
R> # We use a variable not occurring in the dataset
R> badrule <- validator(employees >= 0)
R> confront(retailers, badrule)
\end{Sinput}
\begin{Soutput}
Object of class 'validation'
Call:
    confront(dat = retailers, x = badrule)

Confrontations: 1
With fails    : 0
Warnings      : 0
Errors        : 1
\end{Soutput}
\end{Schunk}
In section \S\ref{sect:confronting} we explain in more detail how to handle
errors or warnings. Here, we return to the original rule set stored in
\code{rules} and summarize the results. 
\begin{Schunk}
\begin{Sinput}
R> check <- confront(retailers, rules, key="id")
R> summary(check)
\end{Sinput}
\begin{Soutput}
   name items passes fails nNA error warning
1    st    60     54     0   6 FALSE   FALSE
2    to    60     56     0   4 FALSE   FALSE
3    or    60     23     1  36 FALSE   FALSE
4 st.cs    60     50     0  10 FALSE   FALSE
5    bl    60     19     4  37 FALSE   FALSE
6    mn     1      1     0   0 FALSE   FALSE
                                     expression
1                         (staff - 0) >= -1e-08
2                      (turnover - 0) >= -1e-08
3                     (other.rev - 0) >= -1e-08
4              !(staff > 0) | (staff.costs > 0)
5 abs(turnover + other.rev - total.rev) < 1e-08
6               mean(profit, na.rm = TRUE) >= 1
\end{Soutput}
\end{Schunk}
The \code{summary} method returns a data frame where each row represents one
validation rule. The second column (items) lists how many results each rule
returned, i.e., how many items were checked with the rule.  Here, the first
five rules are executed for each row in the data set so there are 60 items
(each row is one item). The last rule returns a single value. Calculating this
value involves comparing the mean profit with a positive number, so the
`profit' column is the single item under scrutiny.

The next two columns list how many items passed, or failed the test. The
\code{nNA} column shows how many tests resulted in \code{NA} because one or
more of the data points needed to evaluate the rule were missing. The columns
named \code{error} and \code{warning} indicate whether an error or warning
occurred during evaluation, and the last column shows the actual expression
used to evaluate the rule. Depending on the rule, the original expressions are
manipulated, for example to account for machine rounding errors (rule 1--3, and
5) or to vectorize the statement (rule 4). The choices made in these
conversions can be influenced with parameters that will be detailed in
Subsection~\ref{sect:confrontation}.

The full set of results can be extracted, for example in the form of a data
frame:
\begin{Schunk}
\begin{Sinput}
R> output <- as.data.frame(check)
R> tail(output,3)
\end{Sinput}
\begin{Soutput}
       id name value                                    expression
299 RET59   bl    NA abs(turnover + other.rev - total.rev) < 1e-08
300 RET60   bl    NA abs(turnover + other.rev - total.rev) < 1e-08
301  <NA>   mn  TRUE               mean(profit, na.rm = TRUE) >= 1
\end{Soutput}
\end{Schunk}
The output is a record for each validation on each item, with columns
identifying the item (if possible), the name of the validation rule, the result
and the \proglang{R} expression used in obtaining the result. 

A very short summary of the results can be extracted with \code{all}, which
returns \code{TRUE} only when all tests are passed.
\begin{Schunk}
\begin{Sinput}
R> # passing all checks?
R> all(check)
\end{Sinput}
\begin{Soutput}
[1] FALSE
\end{Soutput}
\begin{Sinput}
R> # ignore missings:
R> all(check, na.rm=TRUE)
\end{Sinput}
\begin{Soutput}
[1] FALSE
\end{Soutput}
\end{Schunk}

\begin{figure}
\centering
\includegraphics[width=0.5\textwidth]{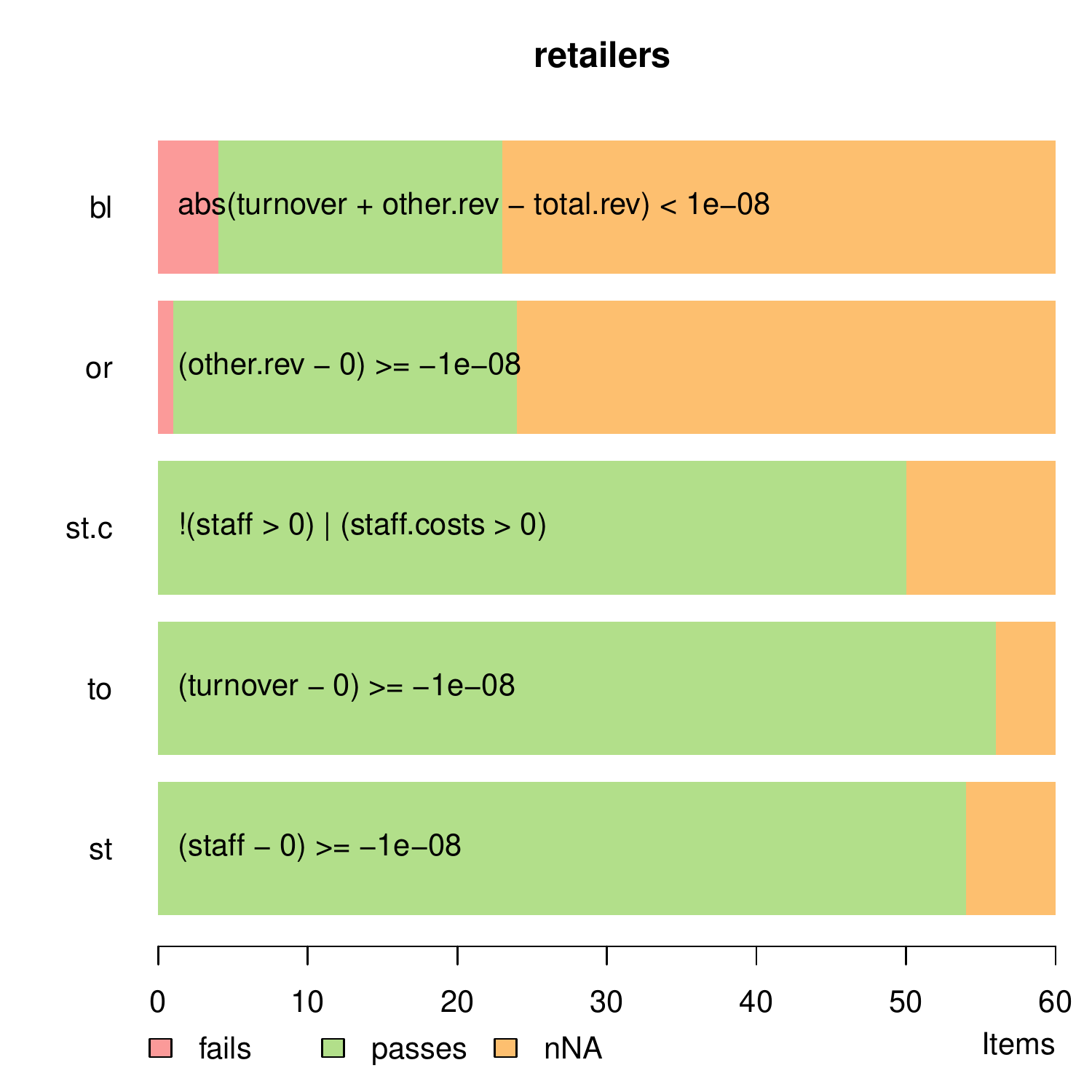}
\caption{Barplot of validation results.}
\label{fig:barplot}
\end{figure}

With \code{plot} we can quickly visualize the results. Here, we only visualize
results of rules 1--5 because these are the rules rules that are applied to
more than one item. The result is shown in Figure~\ref{fig:barplot}.
\begin{Schunk}
\begin{Sinput}
R> plot(check[1:5], main='retailers')
\end{Sinput}
\end{Schunk}

Until now we used \code{validator} to first create a rule set and then
\code{confront}, to confront data with rules. In the following sections we
shall sometimes use \code{check\_that} which combines these steps.
\begin{Schunk}
\begin{Sinput}
R> ## this
R> out <- check_that(retailers, staff >= 0, other.rev >= 0)
R> ## is equivalent to
R> out <- confront(retailers, validator(staff>=0, other.rev >= 0))
\end{Sinput}
\end{Schunk}

\subsection{Domain specific languages}
\label{sect:dsl}
According to \citet{fowler2010domain}, a domain specific language (DSL) is
``\emph{a computer programming language of limited expressiveness focused on a
particular domain}''. DSLs are commonly divided into \emph{standalone} DSLs
that can be run within a specific software and \emph{embedded} DSLs that are
developed as a subset of expressions of an existing language
\citep{gibbons2013functional}. Examples of \emph{standalone} DSLs are the
\proglang{BLAISE} scripting language for survey questionnaire design
\citep{cbs2018blaise} and the relatively new VTL language for data validation
and transformation \citep{sdmx2018vtl}. A well-known example of a DSL in \proglang{R} is
the set of expressions accepted by the \code{subset} function.  Recall that if
\code{d} is a data frame then \code{subset(d, <expression>)} returns the records
in \code{d} satisfying \code{<expression>}. The DSL for \code{subset} is a set
of \proglang{R} expressions that evaluate to a logical vector suitable for filtering
records from \code{d}.

There are several reasons why \proglang{R} \citep{rcore} is a good choice for
hosting an embedded DSL. First, it offers all the necessary technical tools out
of the box: \proglang{R} expressions can be manipulated and investigated just
like any other variable. In \pkg{validate} this is used to check if an
expression passed to \code{validator} can be interpreted as a validation rule.
\proglang{R} also offers functions such as \code{eval}. This function accepts
an unevaluated \proglang{R} expression and a data set and evaluates the
expression using data in the data set. In \pkg{validate} such a construction is
used to execute validation rules for a data set. Second, \proglang{R} and its
packages offer an immense amount of well-defined and well-tested (statistical)
functionality. A DSL that is built in \proglang{R} inherits all this
functionality for free.

\subsection{A domain specific language for data validation}
\label{sect:dslvalidation}
The DSL implemented in \pkg{validate} consists of \proglang{R} expressions that
evaluate to \class{logical}, plus some special expressions that do not evaluate
to \class{logical} but that make certain common data validation tasks easier.
This section is devoted to the expressions that evaluate to a \class{logical}.
Other syntax elements (`syntactic sugar') are discussed in the next section.

To ensure that an expression stored in a \class{validator} object results in a
\class{logical}, it is inspected when the user defines it.  When a user passes
an expression that is not part of our DSL, it is ignored with a warning.
\begin{Schunk}
\begin{Sinput}
R> rules <- validator(x > 0, mean(x))
\end{Sinput}
\end{Schunk}
\begin{Schunk}
\begin{Soutput}
Warning message:
  Invalid syntax detected, the following expressions have been ignored:
[002] mean(x)
\end{Soutput}
\end{Schunk}
Here, the first expression is recognized as part of the DSL because the final
operation to be evaluated (the \code{>} comparison operator) must yield a
\class{logical}. This is not true for the second expression which just computes
a mean. For each expression defined by the user, the \code{validator} function
compares the final operation with a list of allowed final operations.  An
overview of these operations is given in the table below. They can be separated
into unary and binary operators or functions. Let us clarify their use with some
examples.
\begin{table}[h]
\centering
\caption{Basic operations and functions that define a validation rule.}
\label{tab:operations}
\begin{tabular}{ll}
\hline
Type                                & \proglang{R} function or operator\\
\hline
Unary operators or functions        & \code{!, all, any, grepl}, any \code{is.} function.\\
Binary operators or functions       & \code{<, <=, ==, identical, !=, >=, >, \%in\%, if,}\verb" ~"\\
Helper functions                    & \code{is\_unique}, \code{all\_unique}, \code{is\_complete}, \code{all\_complete}\\
\hline
\end{tabular}
\end{table}

\paragraph{Example 1: checking for duplicates.}
The retailers dataset has an \code{id} variable. Requiering it
to be unique can be expressed as follows.
\begin{Schunk}
\begin{Sinput}
R> checks <- check_that(retailers, !any(duplicated(id))) 
R> all(checks)
\end{Sinput}
\begin{Soutput}
[1] TRUE
\end{Soutput}
\end{Schunk}
Or, using one of the helper functions
\begin{Schunk}
\begin{Sinput}
R> all( check_that(retailers, all_unique(id)) )
\end{Sinput}
\begin{Soutput}
[1] TRUE
\end{Soutput}
\end{Schunk}
The advantage of \code{all\_unique} is that it accepts a comma-separated
list of variable names so it is also possible to check for uniqueness of
combinations of values.

\paragraph{Example 2: checking variable type.}
Any function starting with `\code{is.}' is accepted as a data validation
function. Below we demand that \code{turnover} is a numeric variable and 
\code{size} is a categorical variable (\code{factor}).
\begin{Schunk}
\begin{Sinput}
R> checks <- check_that(retailers, is.numeric(turnover), is.factor(size))
R> all(checks)
\end{Sinput}
\begin{Soutput}
[1] TRUE
\end{Soutput}
\end{Schunk}

The use of comparison operators (\code{<},\code{<=},$\ldots$) was demonstrated
in Section~\ref{sect:datavalidation}. The next example demonstrates how more
complex validation rules can be built up by reusing functionality already
present in \proglang{R}. 
\paragraph{Example 3: correlation between variables.}
Suppose we wish to check whether the correlation between two variables
\code{height} and \code{weight} in the \code{women} dataset is larger than 0.5.
\begin{Schunk}
\begin{Sinput}
R> correlation_check <- validator(cor(height,weight) > 0.5)
R> summary( confront(women, correlation_check) )
\end{Sinput}
\begin{Soutput}
  name items passes fails nNA error warning                expression
1   V1     1      1     0   0 FALSE   FALSE cor(height, weight) > 0.5
\end{Soutput}
\end{Schunk}

The unary operators and functions also include \proglang{R}'s \code{grepl}
function which accepts a regular expression and a \class{character} vector, and
returns \code{TRUE} for each element of the \class{character} vector that
matches the regular expression. 
\paragraph{Example 4: pattern matching a text variable.}
The following rule tests if the \code{size} variable always consists of a
string starting with \code{"sc"}, followed by a number.  
\begin{Schunk}
\begin{Sinput}
R> checks <- check_that(retailers, grepl("^sc[0-9]$", size))
R> all(checks)
\end{Sinput}
\begin{Soutput}
[1] TRUE
\end{Soutput}
\end{Schunk}

\paragraph{Example 5: testing a variable against a code list.}
Since \code{size} is a categorical variable, we can use
the binary \code{\%in\%} operator to check the values against
an explicit list of allowed values.
\begin{Schunk}
\begin{Sinput}
R> checks <- check_that(retailers, size %in% c("sc0","sc1","sc2","sc3"))
R> all(checks)
\end{Sinput}
\begin{Soutput}
[1] TRUE
\end{Soutput}
\end{Schunk}

The binary operations `\code{if}' and `\verb"~"' (tilde) have special
interpretations that need to be explained in more detail. The \code{if}
function is interpreted as a logical implication that results in a truth value.
In propositional logic, it is usually denoted with an arrow as in $P\Rightarrow
Q$ ($P$ implies $Q$), where $P$ and $Q$ are propositions that can be either
TRUE or FALSE. The truth table for this operation is given below.
\begin{center}
\begin{tabular}{rr|r}
\multicolumn{1}{c}{$P$}   &  \multicolumn{1}{c}{$Q$}   & \multicolumn{1}{|c}{$P\Rightarrow Q$}\\
\hline
FALSE &  FALSE & TRUE \\
FALSE &  TRUE  & TRUE \\
TRUE  &  FALSE & FALSE\\
TRUE  &  TRUE  & TRUE \\
\end{tabular}
\end{center}
It is a classic result from propositional logic that $P\Rightarrow Q$ is equal
to $\lnot P\lor Q$ [(not $P$) or $Q$], in the sense that both expressions have
the same truth table. This result is exploited by \pkg{validate} by translating
rules of the form \code{if(P) Q} to \code{!(P) | Q}. This translation allows
\proglang{R} to execute the test in a vectorized manner, resulting in a
significant speed gain.

\paragraph{Example 6: conditional rules.}
When a company employs staff, there should be associated staff costs.
\begin{Schunk}
\begin{Sinput}
R> check <- check_that(retailers, if (staff > 0) staff.costs > 0)
R> all(check, na.rm=TRUE)
\end{Sinput}
\begin{Soutput}
[1] TRUE
\end{Soutput}
\begin{Sinput}
R> summary(check)
\end{Sinput}
\begin{Soutput}
  name items passes fails nNA error warning
1   V1    60     50     0  10 FALSE   FALSE
                        expression
1 !(staff > 0) | (staff.costs > 0)
\end{Soutput}
\end{Schunk}
The summary shows that indeed the original expression has been rewritten.
Comparing the rule with the truth table, we see that according to this rule it
is acceptable for a company to employ no staff and have no staff costs; to have
staff costs without employing staff; and to employ staff and have staff costs.
It is not acceptable to employ staff and not have any staff costs. If it is
reasonable to expect staff whenever there are staff costs, a second rule of the
form \code{if ( staff.costs > 0 ) staff > 0} must be defined.

The last operation we discuss here is the \verb"~" (tilde) operator.  This is
used to indicate so-called functional dependencies. A functional dependency
expresses a relation between multiple records in relational data. For example
given two variables \code{street} and \code{postal\_code} we could specify that
`if two records have the same street name, they must have the same postal
code`. In \pkg{validate}, this would be expressed as
\begin{Schunk}
\begin{Sinput}
R> rules <- validator( street ~ postal_code )
\end{Sinput}
\end{Schunk}
If we also have a variable called \code{city}, we could state that if two
records have the same city and street, they must have the same postal code. This
would be expressed as
\begin{Schunk}
\begin{Sinput}
R> rules <- validator( city + street ~ postal_code )
\end{Sinput}
\end{Schunk}
Similarly, we can express the relation `if two records have the same
postal code, they must have the same values for city and street`.
\begin{Schunk}
\begin{Sinput}
R> rules <- validator( postal_code ~ city + street )
\end{Sinput}
\end{Schunk}
Functional dependencies are have been researched extensively in the field of
database theory. Some interesting sources include
\citet{armstrong1974dependency, beeri1984structure, bohannon2007conditional}
and references therein.

\subsection{Syntactic sugar}
\label{sect:sugar}
The term `syntactic sugar' refers to elements of a syntax that are not strictly
necessary but make common tasks easier. The \pkg{validate} DSL includes some
syntactic sugar to refer to the dataset as a whole, to store and reuse
intermediate results, and to apply the same rule to multiple variables. We
discuss these syntax elements with a few examples while an overview is given in
Table~\ref{tab:sugar}.

\paragraph{Syntactic sugar 1: inspecting metadata.} An important class of data
validation involves checking metadata, including the dimensions of a data set
or the presence of certain variables. In the previous section it was shown how
rules can refer to individual variables in a data set, but to test metadata, it
is convenient to have access to the data set as a whole. This is implemented
using `\code{.}'.
\begin{Schunk}
\begin{Sinput}
R> rules <- validator(
+   nrow(.) >= 100
+  ,"Species" %in% names(.) )
\end{Sinput}
\end{Schunk}
Here, the `\code{.}' refers directly to the data set (here, the \code{iris}
data frame) passed to \code{confront}.

\paragraph{Syntactic sugar 2: reuse calculations.}
The \code{:=} operator can be used to store intermediate calculation for reuse.
In the following rule set we check if the fraction of `versicolor' is between
0.25 and 0.50. To do so, we first compute the fraction and reuse that to check
the bounds.
\begin{Schunk}
\begin{Sinput}
R> rules <- validator(
+   fraction := mean(Species == "versicolor")
+   , vc_upr =  fraction >= 0.25
+   , vc_lwr =  fraction <= 0.50)
\end{Sinput}
\end{Schunk}
The \code{:=} operator can be used as if it were the standard assignment
operator in \code{R}.  The main difference is that the right-hand-side of
\code{A := B} is not evaluated immediately. Rather, in any rule after the
definition of \code{A := B} the use of \code{A} is replaced by \code{B}. We can
see this by confronting the rule set from the example with the \code{iris} data
set and inspecting the expressions that have been evaluated.
\begin{Schunk}
\begin{Sinput}
R> as.data.frame( confront(iris, rules) )["expression"]
\end{Sinput}
\begin{Soutput}
                             expression
1 mean(Species == "versicolor") >= 0.25
2  mean(Species == "versicolor") <= 0.5
\end{Soutput}
\end{Schunk}

\begin{table}[t]
\centering
\caption{Syntactic sugar}
\label{tab:sugar}
\begin{tabular}{lp{0.55\textwidth}}
\hline
Syntax  element     & Description \\
\hline
\code{.}            & Refers to the whole dataset, e.g., \code{nrow(.) > 10}\\
\code{:=}           & Store intermediate result, e.g., \code{med := median(x)}\\
\code{var\_group()} & Create a list of variables to be reused in rules, e.g., \code{G := var\_group(X, Y, Z)}\\
\hline
\end{tabular}
\end{table}

\paragraph{Syntactic sugar 3: referring to external data.}
For some data validation purposes it is necessary or convenient to have access
to external reference data.  For example, we may want to retrieve the list of
valid \code{Species} values from an external source. This can be done as
follows.
\begin{Schunk}
\begin{Sinput}
R> codelist <- data.frame(
+      validSpecies =  c("versicolor", "virginica","setosa")
+    , stringsAsFactors=FALSE)
R> rules <- validator(Species %in% ref$validSpecies)
R> summary(confront(iris, rules, ref=codelist))[1:5]
\end{Sinput}
\begin{Soutput}
  name items passes fails nNA
1   V1   150    150     0   0
\end{Soutput}
\end{Schunk}
Note that the reference data is passed via the \code{ref} argument of
\code{confront}. There are several ways to pass reference data in this way,
including data frames, lists, or environments.

\paragraph{Syntactic sugar 4: same rules, multiple variables.}
When many variables need to satisfy a similar set of rules, rule definition can
become cumbersome. A common example occurs when many variables are measured on
the same scale such as $[0,1]$ or $\{1,2,3,4,5\}$, and all these ranges need to
be checked. Such cases are facilitated with the concept of `variable groups'.
This is essentially a list of variable names that can be reused inside a rule.

Explicitly, the following set of rules (defining that all of 
\code{x}, \code{y}, and \code{z} are between 0 and 1)
\begin{Schunk}
\begin{Sinput}
R> rules <- validator(x >= 0, y >= 0, z >= 0, x <= 1, y <= 1, z <= 1)
\end{Sinput}
\end{Schunk}
is equivalent to
\begin{Schunk}
\begin{Sinput}
R> rules <- validator( G := var_group(x, y, z), G >= 0, G <= 1 )
\end{Sinput}
\end{Schunk}
The rule sets are expanded when confronted with data. If multiple variable
groups are used in a single validation rule, the expansion is based on the
Cartesian product of the variables defined in the variable groups.

\subsection{Validator objects}
\label{sect:validator}
Every data validation rule expresses an assumption about a data set. When the
number of such validation rules grows, it becomes desirable to be able to
manage them with CRUD (create, read, update, delete) operations, to describe,
analyze, and manipulate them. There are also benefits in having access to rule
metadata, such as a rule name or description. Such information can then be
included in automatically generated data quality reports or dashboards. The
purpose of \class{validator} objects is to import, manipulate, and export
validation rules and their metadata. In this section we demonstrate how to
manipulate and investigate validation rules and their metadata, in the next
section we focus on import and export of rule sets.

Objects of class \class{validator} behave much like an \proglang{R}
\code{list}.  They can be subsetted with the square bracket operator, and a
single element can be extracted using double square brackets.
\begin{Schunk}
\begin{Sinput}
R> rules <- validator(minht = height >= 40, maxht = height <= 95)
R> # Subsetting returns a 'validator' object.
R> rules[2]
\end{Sinput}
\begin{Soutput}
Object of class 'validator' with 1 elements:
 maxht: height <= 95
Rules are evaluated using locally defined options
\end{Soutput}
\end{Schunk}
A single element is an object of class \class{rule}. It holds an expression and
some metadata.
\begin{Schunk}
\begin{Sinput}
R> rules[[1]]
\end{Sinput}
\begin{Soutput}
Object of class rule.
 expr       : height >= 40 
 name       : minht 
 label      :  
 description:  
 origin     : command-line 
 created    : 2019-12-20 11:53:49
 meta       : language<chr>, severity<chr>
\end{Soutput}
\end{Schunk}
Here, \code{label} and \code{description} are short and long descriptions of
the rule respectively. These have to be defined by the user. The fields
\code{origin} and \code{created} are automatically filled when the rules are
defined. The \code{meta} field allows the user to add extra metadata elements.

Metadata can be extracted and defined in ways that should be familiar to
\proglang{R} users. For example, setting labels is done as follows.
\begin{Schunk}
\begin{Sinput}
R> label(rules) <- c("least height", "largest height")
R> rules
\end{Sinput}
\begin{Soutput}
Object of class 'validator' with 2 elements:
 minht [least height]  : height >= 40
 maxht [largest height]: height <= 95
\end{Soutput}
\end{Schunk}
For each field there is a get and set function that works similar to the
\code{names} function of \proglang{R}. The \code{meta} field can be manipulated
with the \code{meta} function, which works similar to \proglang{R}'s
\code{attr} function. An overview of functions for manipulating
\class{validator} instances is given in Table~\ref{tab:validator}.
\begin{table}[t]
  \centering
  \caption{An overview of functions to investigate or manipulate \class{validator}
   objects.}
  \label{tab:validator}
  \begin{tabular}{ll}
    \hline
    Function & description\\
    \hline
    \code{summary}        & Summarize properties\\
    \code{plot}           & Matrix plot showing what variables occur in each rule\\
    \code{`[`}            & Create new \class{validator} with subset of rules \\
    \code{`[[`}           & Extract a single rule with its metadata\\
    \code{length}         & Number of rules\\
    \code{as.data.frame}  & Store rules (as \code{character}) and metadata in data frame\\
    \code{variables}      & Which variables occur in a \class{validator}?\\ 
    \code{names, names<-} & get, set rule names\\
    \code{label, label<-} & get, set rule labels\\
    \code{description, description<-} & get, set rule description\\
    \code{origin, origin<-} & get, set rule origin\\
    \code{created, created<-} & get, set rule timestamp\\
    \code{meta, meta<-}   & get, set generic metadata\\
    \code{+}              & combine two \code{validator} instances\\
  \hline
  \end{tabular}
\end{table}

For larger rule sets it is interesting to query which variables are covered and
by which rules. This is done with \code{variables}, which returns an overview
of variable coverage by the rule set.
\begin{Schunk}
\begin{Sinput}
R> variables(rules)
\end{Sinput}
\begin{Soutput}
[1] "height"
\end{Soutput}
\begin{Sinput}
R> variables(rules, as='matrix')
\end{Sinput}
\begin{Soutput}
       variable
rule    height
  minht   TRUE
  maxht   TRUE
\end{Soutput}
\end{Schunk}
One use is to automatically detect whether all variables in 
a data set are covered in by a rule set.
\begin{Schunk}
\begin{Sinput}
R> all(names(women) %in% variables(rules))
\end{Sinput}
\begin{Soutput}
[1] FALSE
\end{Soutput}
\begin{Sinput}
R> names(women)[!names(women) %in% variables(rules)]
\end{Sinput}
\begin{Soutput}
[1] "weight"
\end{Soutput}
\end{Schunk}

There are other interesting questions to be asked about rule sets, for example
whether there are any redundancies or even contradictions. For this more
advanced type of functionality the reader is referred to the complementary
\pkg{validatetools} package \citep{jonge2018validatetools}.

\subsection{Import and export of rule sets}
\label{sect:importexport}
Validation rules can be defined in several unstructured and structured
formats. The unstructured formats include definition on the command line and
definition via an unstructured text file (source code). Structured formats
include data frames or \proglang{YAML} files \citep{yaml2015} where each rule
can be endowed with metadata. Here we discuss some advantages and disadvantages
of the methods for storing and exporting data validation rule sets (see also
Table~\ref{tab:impexp}).
\begin{table}[t]
\centering
\caption{Import and export of validation rules.}
\label{tab:impexp}
\begin{tabular}{ll}
\hline
Function & Description\\
\hline
\code{validator(...)} & Define rules on the command-line\\
\code{validator(.file=)} & Read rules from file\\
\code{validator(.data=)} & Read rules from data frame\\
\code{export\_yaml}      & Export to \proglang{YAML} format\\
\code{as.data.frame}     & Transform \class{validator} to \class{data.frame} (lossy)\\
\hline
\end{tabular}
\end{table}

A rule set and its metadata can be exported to data frame using
\code{as.data.frame} as usual in \proglang{R}. Reading from data frame is done
using \code{validator(.data=)}. Import from and export to data frame is
especially useful in situations where rule sets are stored in a centralized
repository that is implemented as a data base. A disadvantage of using data
frames is that metadata that pertains to the rule set as a whole will be lost
when exporting to data frame. In particular, option settings particular to a
rule set are not retained when exporting to data frame. For extensive rule sets
an approach using structured or unstructured text files, or a combination
thereof may better suit the needs.

Text files offer the most flexible way of defining rule sets in \pkg{validate}.
Indeed, there are benefits in treating a rule set as a type of source code that
can be managed under version control and provided with comments. The
\pkg{validate} package also supports advanced features such as file inclusion,
where one rule file can refer to another rule file. We will demonstrate this in
the following example, which also shows how structured and unstructured formats
may be used together.

For this example we imagine the following use case. We have a questionnaire
that consists of two parts: a general part that is sent to all participants of
a study, and a specific part of which the contents depends on the participant
type. Here, the general part is related to retailers, while the specific part
is related to the subdomain of supermarkets. The idea is that we maintain two
files: one with general rules, that is to be reused accross different
subdomains and one with specific rules for supermarkets.

Figure~\ref{fig:yaml} shows the general rules in \proglang{YAML} format. At
the top, there is a section with general options pertaining to the rule set as
a whole. Next, all the rules and their metadata are defined one by one. The
rules are stored in a file named \code{general\_rules.yml}. The specific rules
are stored in a simpler text file called \code{rules.txt} with the following
content.
\begin{Code}
  ---
  include:
    - general_rules.yml
  ---

  # a reasonable profit
  profit/total.rev <= 0.6

  # We expect that the supermarket sector
  # is profitable on average
  mean(profit) >= 1
\end{Code}
At the top there is an optional block, demarcated with dashes, where we can
define properties for the whole rule set or as in this case indicate that the
file \code{general\_rules.yml} must be read before the current file. Both files
can now be read as follows.
\begin{Schunk}
\begin{Sinput}
R> rules <- validator(.file="rules.txt")
R> rules
\end{Sinput}
\begin{Soutput}
Object of class 'validator' with 5 elements:
 G1 [nonnegative staff] : staff >= 0
 G2 [nonnegative income]: turnover >= 0
 G3 [Balance check]     : profit + total.costs == total.rev
 V1                     : profit/total.rev <= 0.6
 V2                     : mean(profit) >= 1
\end{Soutput}
\end{Schunk}
\begin{figure}[!t]
\hrule
\begin{Code}

---
options:
  raise: none
  lin.eq.eps: 1.0e-08
  lin.ineq.eps: 1.0e-08
---
rules:
- expr: staff >= 0
  name: 'G1'
  label: 'nonnegative staff'
  description: |
    'Staff numbers cannot be negative'
  created: 2018-06-05 14:44:06
  origin: 
  meta: []
- expr: turnover >= 0
  name: 'G2'
  label: 'nonnegative income'
  description: | 
    'Income cannot be negative (unlike in the
     definition of the tax office)'
  created: 2018-06-05 14:44:06
  origin: 
  meta: []
- expr: profit + total.costs == total.rev
  name: 'G3'
  label: 'Balance check'
  description: |
    'Economic profit is defined as the
     total revenue diminished with the 
     total costs.'
  created: 2018-06-05 14:44:06
  origin: 
  meta: []
\end{Code}
\hrule
\caption{Contents of the file \code{general\_rules.yml}. Several global options
and metadata fields have been filled in.}
\label{fig:yaml}
\end{figure}

A user is able to trace the origins of each rule interactively since the
\class{validator} remembers where each rule came from.
\begin{Schunk}
\begin{Sinput}
R> origin(rules)
\end{Sinput}
\begin{Soutput}
                   G1                    G2                    G3 
"./general_rules.yml" "./general_rules.yml" "./general_rules.yml" 
                   V1                    V2 
          "rules.txt"           "rules.txt" 
\end{Soutput}
\end{Schunk}

Finally we point out that file inclusion works recursively so included files
can include other files. Mistakes such as cyclic inclusion are detected and
reported when they occur. A complete description can be found in a dedicated
vignette that is included with the package.

\subsection{Confronting data with rules}
\label{sect:confronting}
At the beginning of Section~\ref{sect:validate} we have seen how data can be
checked against a rule set using \code{confront}.  It was also highlighted that
several default choices were made, for instance how machine rounding and
exceptions (errors, warnings) are handled, and the fact that missing values in
data lead to missing data validation results. In this section we demonstrate
how those options can be manipulated both globally and locally.

\begin{table}[t]
\centering
\caption{Options for confronting data with rules. Default values in brackets.}
\label{tab:options}
\begin{tabular}{lll}
\hline
Option            & Values                          & Description \\
\hline
\code{na.value}   & \code{[NA], TRUE, FALSE}        & Value when rule results in \code{NA}\\
\code{raise}      & \code{["none"], "error", "all"} & What exceptions to raise\\
\code{lin.eq.eps} & \code{[1e-8]}; positive number  & Tolerance for checking equalities\\ 
\code{lin.ineq.eps} & \code{[1e-8]}; positive number& Tolerance for checking inequalities\\ 
\hline
\end{tabular}
\end{table}

There are three places where options can be set, with three different ranges of
influence. The first place is while calling \code{confront}.  In this case the
options only pertain to the current call. For example, we can count missing
validation results as \code{FALSE} (fail) as follows (we suppress the last
column of the summary for brevity)
\begin{Schunk}
\begin{Sinput}
R> data("retailers")
R> rules <- validator(turnover >= 0, turnover + other.rev == total.rev)
R> summary( confront(retailers, rules) )[-8]
\end{Sinput}
\begin{Soutput}
  name items passes fails nNA error warning
1   V1    60     56     0   4 FALSE   FALSE
2   V2    60     19     4  37 FALSE   FALSE
\end{Soutput}
\begin{Sinput}
R> summary( confront(retailers, rules, na.value=FALSE) )[-8]
\end{Sinput}
\begin{Soutput}
  name items passes fails nNA error warning
1   V1    60     56     4   0 FALSE   FALSE
2   V2    60     19    41   0 FALSE   FALSE
\end{Soutput}
\end{Schunk}
The second place to control options is by setting options for a specific
\class{validator} instance. This is done with the \code{voptions} function. For
example, if we know that all our variables are integer we can ignore the
possibility of spurious fails caused by machine rounding in linear equalities
and linear inequalities (we ignore columns 6 and 7 in the output for brevity).
\begin{Schunk}
\begin{Sinput}
R> voptions(rules, lin.eq.eps=0, lin.ineq.eps=0)
R> rules
\end{Sinput}
\begin{Soutput}
Object of class 'validator' with 2 elements:
 V1: turnover >= 0
 V2: turnover + other.rev == total.rev
Rules are evaluated using locally defined options
\end{Soutput}
\begin{Sinput}
R> summary( confront(retailers, rules) )[-(6:7)]
\end{Sinput}
\begin{Soutput}
  name items passes fails nNA                        expression
1   V1    60     56     0   4                     turnover >= 0
2   V2    60     19     4  37 turnover + other.rev == total.rev
\end{Soutput}
\end{Schunk}
The expressions in the \code{expression} column are now exactly equal to the
user-defined rules and there is no tolerance that allows for some machine
rounding.

The third and final place where options can be set is globally. This is done by
calling \code{voptions} with \code{option = value} pairs.  This will affect all
\code{validator} objects, except those that already have their options adjusted. 
Recall that by default, errors are caught and stored. 
\begin{Schunk}
\begin{Sinput}
R> data("retailers")
R> out <- check_that(retailers, employees >= 0) # the error is caught.
\end{Sinput}
\end{Schunk}
However, we can set \code{raise="error"} so execution stops when an error is
raised.  
\begin{Schunk}
\begin{Sinput}
R> voptions(raise = "error") # set global option.
R> out <- check_that(retailers, employees >= 0)
Error in fun(...) : object 'employees' not found
\end{Sinput}
\end{Schunk}
Raising errors or warnings immediately rather then just registering them can be
useful for example when developing or debugging a large rule set.

To summarize, any option listed in Table~\ref{tab:options} can be set at three
levels: globally, for individual \class{validator} objects, and during a single
call to \class{confront}.

\subsection{Validation objects and analyzing results}
\label{sect:confrontation}
The return value of \code{confront} is a \class{validation} object.  It holds
the data validation results and some information on the data validation
procedure. This information can be extracted and summarized with the functions
listed in Table~\ref{tab:confrontation}. Below we demonstrate some of them and
point out some issues related to the dimensionality of data validation results.

Like \class{validator} objects, \class{validation} objects can be subsetted
by rule using single square brackets. 
\begin{Schunk}
\begin{Sinput}
R> rules <- validator(other.rev >= 0, turnover >= 0,
+   turnover + other.rev == total.rev)
R> check <- confront(retailers, rules)
R> summary(check[1:2])
\end{Sinput}
\begin{Soutput}
  name items passes fails nNA error warning                expression
1   V1    60     23     1  36 FALSE   FALSE (other.rev - 0) >= -1e-08
2   V2    60     56     0   4 FALSE   FALSE  (turnover - 0) >= -1e-08
\end{Soutput}
\end{Schunk}
Using \code{values} all results can be gathered in an array.
\begin{Schunk}
\begin{Sinput}
R> head(values(check), n=3)
\end{Sinput}
\begin{Soutput}
        V1   V2    V3
[1,]    NA   NA    NA
[2,]    NA TRUE    NA
[3,] FALSE TRUE FALSE
\end{Soutput}
\end{Schunk}

By aggregating and sorting results by rule or by record, interesting
information can be gathered on which records are the `worst violators', or
which rules are violated most often. 
\begin{Schunk}
\begin{Sinput}
R> sort(check, by="rule")
\end{Sinput}
\begin{Soutput}
   npass nfail nNA  rel.pass   rel.fail     rel.NA
V3    19     4  37 0.3166667 0.06666667 0.61666667
V1    23     1  36 0.3833333 0.01666667 0.60000000
V2    56     0   4 0.9333333 0.00000000 0.06666667
\end{Soutput}
\end{Schunk}
Here, we aggregated results by rule and sorted the totals by increasing number
of passes per rule. In this case the balance restriction (rule \code{V3})
is passed the least number of times.
\begin{table}
  \centering
  \caption{Investigating \class{confrontation}
  objects.}
  \label{tab:confrontation}
  \begin{tabular}{ll}
    \hline
    Function & description\\
    \hline
    \code{summary}              & Summarize results per rule\\
    \code{all}                  & Check if all validations result in \yay\\
    \code{any}                  & Check if any validation resulted in \nay\\  
    \code{as.data.frame}        & Gather results in a data frame\\
    \code{values}               & Gather results in a (list of) array(s)\\
    \code{aggregate}            & Aggregate results by record or by rule\\
    \code{sort}                 & As \code{aggregate}, but with sorted results\\
    \code{plot}, \code{barplot} & Create barplot(s) of results\\
    \code{errors}               & List of error signals\\
    \code{warnings}             & List of warning signals\\
    \code{`[`}                  & Select subset of confrontations (by rule) \\
    \code{length}               & Number of rules evaluated.\\
  \hline
  \end{tabular}
\end{table}

One issue that hampers analysis of data validation results is that the outcome
of different validation rules may have different dimensions. As an example
consider the following rules for the \code{retailers} dataset.
\begin{Schunk}
\begin{Sinput}
R> rules <- validator(staff >= 0, turnover >= 0, mean(profit) >= 1)
R> check <- confront(retailers, rules)
\end{Sinput}
\end{Schunk}
The first two rules evaluate to sixty results: one for each record in the
\code{retailers} data frame. The third rule evaluates to a single result for
the whole data set. This means that \code{values} cannot meaningfully combine
all results in a single array. For this reason \code{values} works much like
\proglang{R}'s native \code{sapply} function. An array is returned when all
results fit in a single array. If not, a list is returned with an array
for each subset of results that fit together in an array.
\begin{Schunk}
\begin{Sinput}
R> c( class(values(check[1:2])), class(values(check)) )
\end{Sinput}
\begin{Soutput}
[1] "matrix" "list"  
\end{Soutput}
\end{Schunk}
The default behavior can be controlled with a \code{simplify} argument. It has
default value \yay, again similar to \code{sapply}.

\section{Object model and implementation}
\label{sect:implementation}
The \pkg{validate} package is designed with extensibility in mind. The main
objects and methods are based on the \code{S4} system (see e.g.,
\citet{chambers2016extending}) since this is the most flexible object system
currently available in base \proglang{R}. In particular, the fact that
\code{S4} supports multiple dispatch was an important driver to choose it over
\code{S3}. This allows us to easily add \code{confront} methods for different
types of rules or different data types.

\begin{figure}[!t]
\centering
\includegraphics[width=0.9\textwidth]{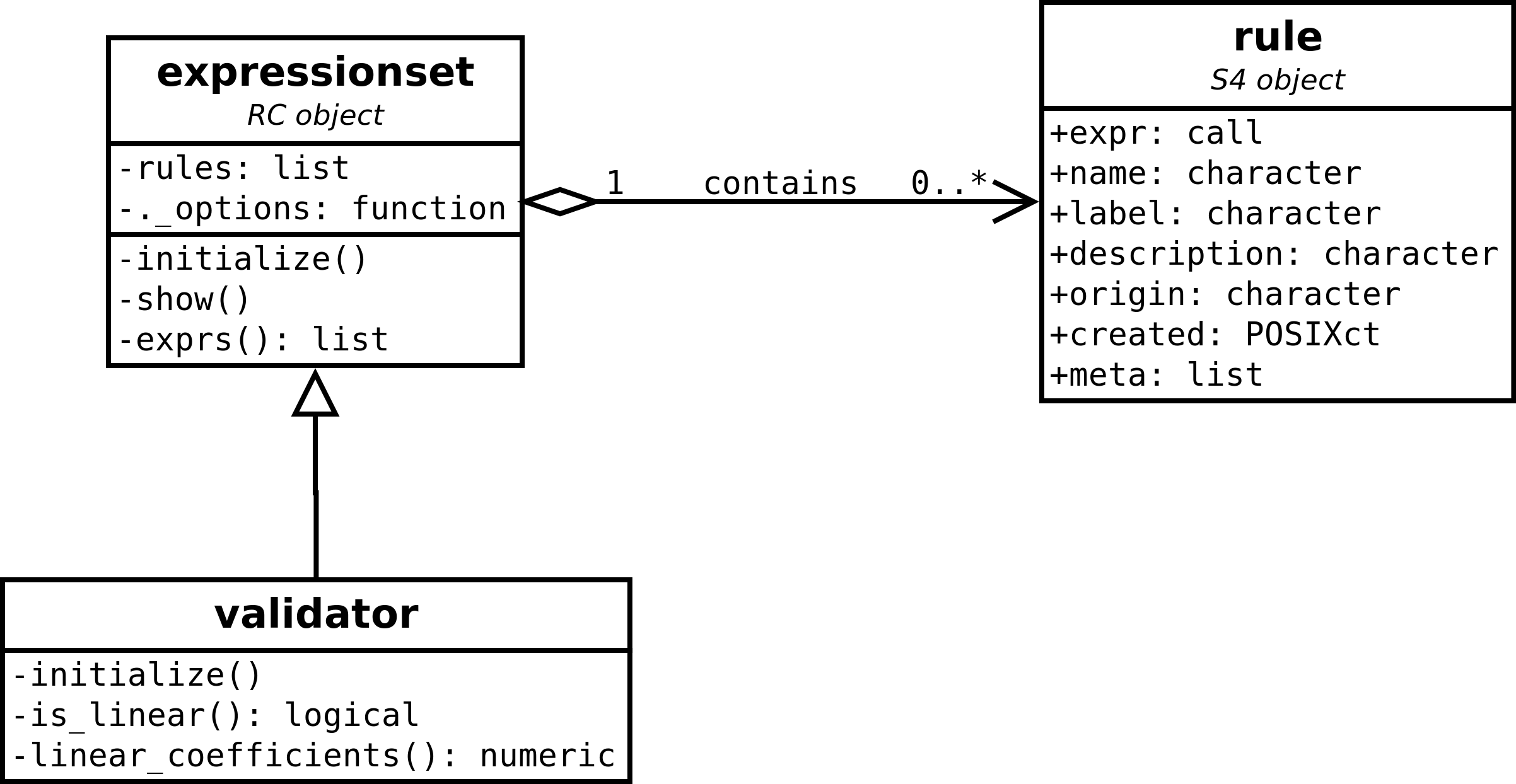}
\caption{UML diagram demonstrating part of the \class{validator} object model.
See \citet{rumbaugh2004UML} for a description of UML.}
\label{fig:validator}
\end{figure}

\begin{figure}[!t]
\centering
\includegraphics[width=0.6\textwidth]{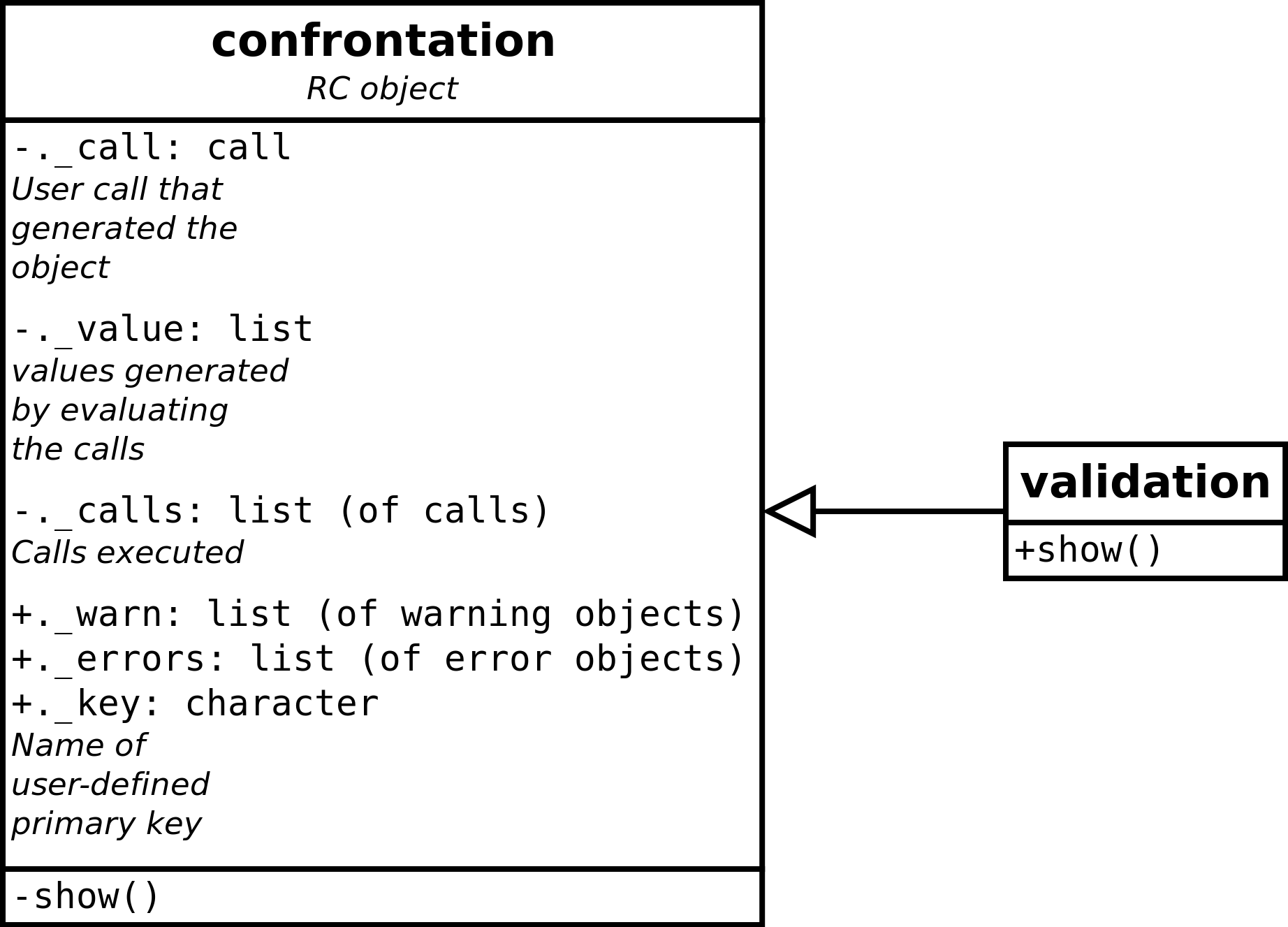}
\caption{UML diagram showing contents of the \class{confrontation} and
\class{validation} object classes.}
\label{fig:confrontation}
\end{figure}

Figures~\ref{fig:validator} and \ref{fig:confrontation} give a high-level
overview of the object model used in \pkg{validate} in UML diagrams.  The
central object for storing expressions is a (not user-visible) class called
\class{expressionset}. An expressionset contains an option settings manager and
a list of \class{rule} objects. A \class{rule} is an \proglang{R} expression
endowed with with some standard metadata.  The object type that users see and
manipulate directly is called \class{validator}. This is a specialization of
\class{expressionset} with some methods that only apply to data validation
rules.  For example, the constructor of \class{validator} checks whether the
imported rules fall within the Domain Specific Language described in
Subsection~\ref{sect:dslvalidation}. The constructor of \class{expressionset}
does not perform such checks. There are also a few methods for extracting
properties of validation rules. These are currently not publicly
documented and should be avoided by users. The reason they are implemented as
methods of \code{validator} objects is that it allows us to reuse them in
packages depending on \pkg{validate} (for example \pkg{validatetools}).

The user-visible return value of \code{confrontation} is an object of class
\class{validation} (Figure~\ref{fig:confrontation}). This is again a
specialization of a more general object called \class{confrontation}.  A
\class{confrontation} object contains the call that created it, the evaluated
expressions and their results. Currently the specialization in
\class{validation} is limited to the \code{show} method.

\newpage{}
\section{Demo: cleaning up a small data set}
\label{sect:demo}
In this Section we demonstrate how \pkg{validate} can be integrated in a data
cleaning workflow. We also introduce two new functions called \code{cells} and
\code{compare}, that allow for comparing two or more versions of the same
dataset.

The purpose in this example is to define a set of rules for the
\code{retailers} data set and to process it step by step until all fields are
completed and all rules are satisfied. Moreover, we wish to monitor changes in
the quality of the dataset as it gets processed.  To this end we use the
following \proglang{R} packages (to be discussed and referenced below).
\begin{Schunk}
\begin{Sinput}
R> library("validate")
R> library("dcmodify")
R> library("errorlocate")
R> library("simputation")
R> library("rspa")
\end{Sinput}
\end{Schunk}
The task is to clean up measured attributes of the \code{retailers} dataset.
This excludes the first two columns (representing size class and sample
inclusion probability) so we first create a working copy for the example.
\begin{Schunk}
\begin{Sinput}
R> data("retailers")
R> dat_raw <- retailers[-(1:2)]
\end{Sinput}
\end{Schunk}
We have defined a set of 18 validation
rules for this data set in a text file that is available and documented in the
supplementary materials. The rules include account balances, non-negativity
constraints and sanity constraints on ratios between variables. For brevity
only a few examples are displayed below. We relax the condition on equalities
somewhat to accommodate a method that will be introduced further on.
\begin{Schunk}
\begin{Sinput}
R> rules <- validator(.file="rules.R")
R> voptions(rules, lin.eq.eps=0.01)
R> rules[c(1,4,11)] # show a few examples
\end{Sinput}
\begin{Soutput}
Object of class 'validator' with 3 elements:
 V01: staff >= 0
 V04: turnover + other.rev == total.rev
 V11: profit <= 0.6 * turnover
Rules are evaluated using locally defined options
\end{Soutput}
\end{Schunk}

The \code{retailers} set consists of staff numbers and financial variables of
60 supermarkets.  Respondents have been asked to submit financial amounts in
multiples of EUR~1000, but in some cases they are submitted in Euro. For
example, this is obviously the case in record 15.
\begin{Schunk}
\begin{Sinput}
R> dat_raw[15,c('staff','turnover','staff.costs')]
\end{Sinput}
\begin{Soutput}
   staff turnover staff.costs
15     3    80000       40000
\end{Soutput}
\end{Schunk}
Taking this record at face value would be equal to believing that a retailer
with a staff of 3 generates 80 million Euros turnover and pays 40 million to
three employees. If we assume that \code{staff} is a reliable variable, such
errors can often be detected and repaired using rule-based processing.  For
example, \emph{\code{IF} the ratio between \code{turnover} and \code{staff}
exceeds 500 \code{THEN} divide \code{turnover} by 1000}. The \pkg{dcmodify}
package \citep{loo2018dcmodify} allows users to formulate such rules and apply
them to data, much like the way validation rules are defined and applied in the
\pkg{validate} package. We defined
12 of such rules, a few of
which are printed here for brevity. The complete set is available in the
supplementary materials.
\begin{Schunk}
\begin{Sinput}
R> modifiers <- dcmodify::modifier(.file="modifiers.R")
R> modifiers[c(1,4,11)]
\end{Sinput}
\begin{Soutput}
Object of class modifier with 3 elements:
M01: 
  if (other.rev >= 500 * vat) other.rev <- other.rev/1000

M04: 
  if (abs(profit) >= 100 * vat) profit <- profit/1000

M11: 
  if (staff.costs >= 500 * staff) staff.costs <- staff.costs/1000
\end{Soutput}
\end{Schunk}
The first and fourth statement compare a financial variable with the value of
the Value-Added-Tax amount (which is considered reliable).  Just like in
\pkg{validate}, the if-statement is executed for each record of the data set.
In cases where the condition evaluates to \yay{} the variable is to be
replaced with the same value divided by one thousand. We can apply the rules to
the dataset as follows.
\begin{Schunk}
\begin{Sinput}
R> dat_mod <- dcmodify::modify(dat_raw, modifiers)
\end{Sinput}
\end{Schunk}

\begin{figure}[t]
\centering
\includegraphics[width=0.7\textwidth]{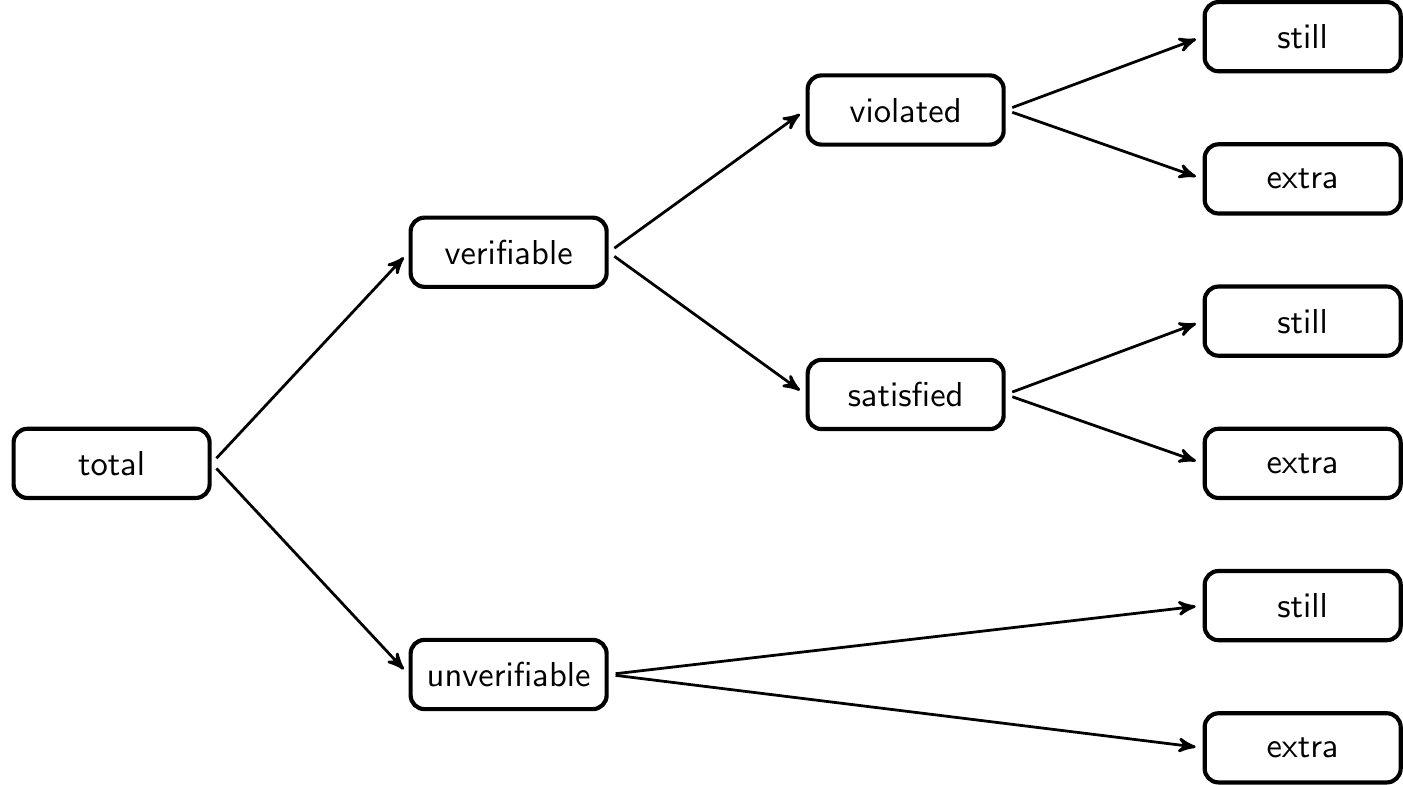}
\caption{Partitioning the result of comparing validation results for two
versions of a data set. The number of cells in a parent node is the sum of the
number of cells in its child nodes (reproduced from
\citet{loo2018statistical}).}
\label{fig:ruletransitions}
\end{figure}

We are now interested in the effect of this step by comparing the validation
results before and after executing these data modifying rules.
Figure~\ref{fig:ruletransitions} shows how we can decompose the validation
results. The total number of validation results can be separated into those
that are verifiable (i.e., yielding \yay{} or \nay{}) and those that are not
(yielding \NA{}). The values that are verifiable can be further separated into
those that are \yay{} or \nay{}, and for each one we can check whether they had
a similar value in the previous version of the data set (still) or if it has
changed (extra). Similarly, we can check for unverifiable validations in the
current version whether they were also unverifiable in the previous version
(still) or not (extra).

The \pkg{validate} package exports a function that computes the number of
transitions for each instance shown in Figure~\ref{fig:ruletransitions}. The
\code{compare} function accepts a \class{validator} and an arbitrary number of
data sets that can be compared sequentially or against the first data set.
\begin{Schunk}
\begin{Sinput}
R> compare(rules, raw = dat_raw, modified=dat_mod)
\end{Sinput}
\begin{Soutput}
Object of class validatorComparison:

   compare(x = rules, raw = dat_raw, modified = dat_mod)

                    Version
Status                raw modified
  validations        1080     1080
  verifiable          789      789
  unverifiable        291      291
  still_unverifiable  291      291
  new_unverifiable      0        0
  satisfied           746      755
  still_satisfied     746      741
  new_satisfied         0       14
  violated             43       34
  still_violated       43       29
  new_violated          0        5
\end{Soutput}
\end{Schunk}
We see that out of 1080 validations ($=$ 60 records times $18$ rules), 746 are
satisfied in the raw data while 755 are satisfied in the modified data. This
seems an improvement, but a closer look reveals that actually 14 cases were
resolved while 5 new violations were introduced. So the assumptions laid down
in the modifying rules may have been too strong in some cases. Or, it is
possible that one or more records had all financial variables a factor of
thousand too high (thus consistent in terms of balance checks) but not all
variables were altered, hence introducing inconsistency. For this example we
will not dive deeper into this but instead move on to the next cleaning step. 

We now clean up the data in three steps. First, knowing what rules each record
violates, we need to figure out which fields to alter so that all rules can
ultimately be satisfied. This procedure is called `error localization'. Next,
we impute the fields that are empty or deemed erroneous during the error
localization procedure. Since our imputation method will not take validation
rules into account, imputation will generally not yield a dataset that
satisfies all validation rules\footnote{For some work on imputing data under
restrictions see \citet{waal2017imputation, vink2015restrictive} and
\citet{tempelman2007imputation} and references therein.}. So in a last step we
adjust the imputed values minimally (in a sense to be defined) such that all
constraints are satisfied.

For the first step we apply the principle of \citet{fellegi1976systematic}. In
this approach, it is assumed that errors are both rare and independently
distributed over the variables. The idea is to find for each record, the
minimal (weighted) number of fields that can be emptied such that imputation
consistent with the rules is possible in principle. The chief difficulty is
that variables typically occur in more than one rule, so changing a variable to
fix one rule, may cause violation of another rule.  There are several numerical
approaches to solving this \citep{waal2011handbook}. The \pkg{errorlocate}
package implements a mixed integer programming (MIP) approach that both
localizes and removes erroneous values. In the following step we set higher
weights for \code{staff} and \code{vat}, so \pkg{errorlocate} will not replace
them unless it cannot be avoided. This means we judge variables with heigher
weight to be of higher quality.
\begin{Schunk}
\begin{Sinput}
R> weights = setNames(rep(1, ncol(dat_raw)), names(dat_raw))
R> weights[c("staff","vat")] <- 10
R> dat_el <- errorlocate::replace_errors(dat_mod, rules, weight = weights)
R> colSums( summary(confront(dat_el, rules))[3:5] )
\end{Sinput}
\begin{Soutput}
passes  fails    nNA 
   677      0    403 
\end{Soutput}
\end{Schunk}
The output of \code{confront} confirms that there are no more violated rules.
Of course many cannot be checked anymore because a number of cells either were
empty or have been emptied by the error localization routine.

For the imputation step we use the \pkg{simputation} package
\citep{loo2017simputation} to estimate the missing values using classification
and regression trees (CART, \citet{breiman1984cart}).
\begin{Schunk}
\begin{Sinput}
R> dat_imp <- simputation::impute_cart(dat_el, . ~ .)
R> colSums(summary(confront(dat_imp, rules))[3:5])
\end{Sinput}
\begin{Soutput}
passes  fails    nNA 
  1003     77      0 
\end{Soutput}
\end{Schunk}
As expected, we see many fails but no more missing values.

Third and finaly, we adjust the imputed values to fit the constraints. For
this, we use the \pkg{rspa} package \citep{loo2018rspa} which implements the
successive projection algorithm. The idea is to minimize the weighted Euclidean
distance between the current imputed values and adjusted values while
satisfying all restrictions (this algorithm is limited to systems of linear
equalities and inequalities). To account for differences of scale between the
variables, we use the reciprocal of the current values as weight. It was shown
by \citet{zhang2015optimal} that this preserves the ratio between the original
values to first order.
\begin{Schunk}
\begin{Sinput}
R> # Compute weights for weighted Euclidean distance
R> W <- t(t(dat_imp)) # convert to numeric matrix
R> W <- 1/(abs(W)+1)  # compute  weights
R> W <- W/rowSums(W)  # normalize by row
R> # adjust the imputed values.
R> dat_adj <- rspa::match_restrictions(dat_imp, rules
+    , adjust = is.na(dat_el), weight=W, maxiter=1e4)
\end{Sinput}
\end{Schunk}
We can now verify that all rules are satisfied.
\begin{Schunk}
\begin{Sinput}
R> all(confront(dat_adj, rules))
\end{Sinput}
\begin{Soutput}
[1] TRUE
\end{Soutput}
\end{Schunk}

\begin{figure}
\centering
\includegraphics[width=0.5\textwidth]{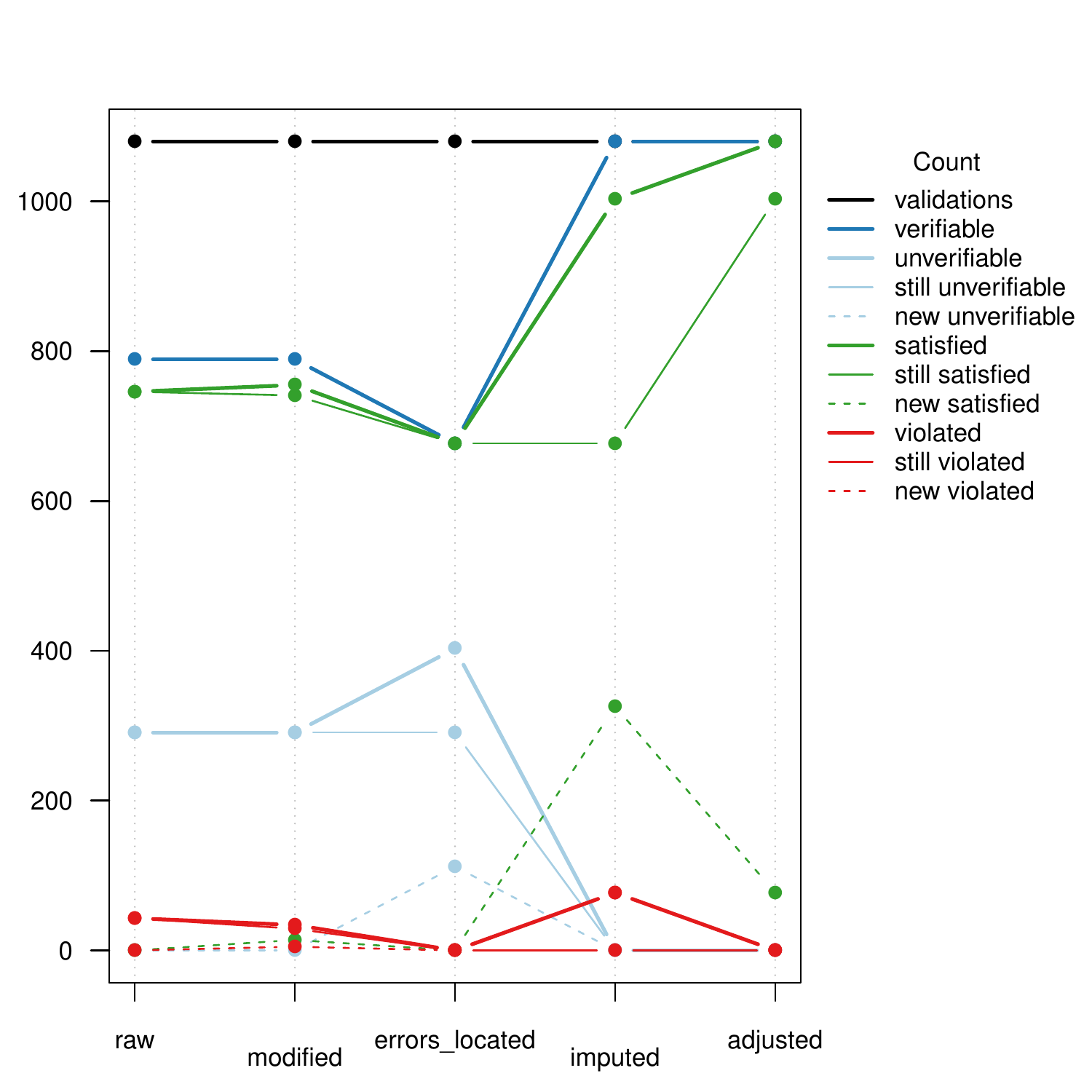}%
\includegraphics[width=0.5\textwidth]{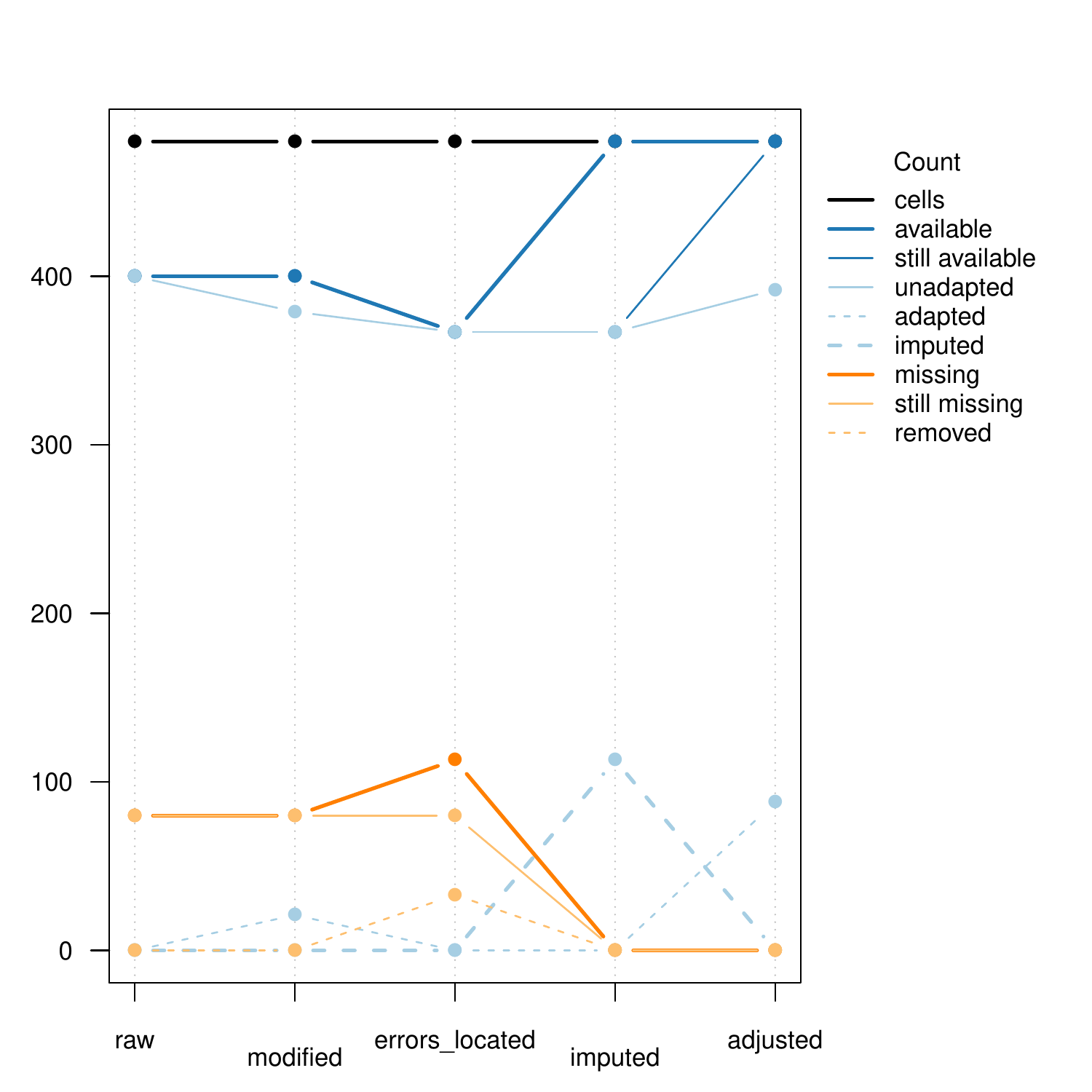}
\caption{Left: following the performance of a dataset with respect to rule satisfaction
as it gets processed step by step. Right: follow changes in cell values as a data set
gets processed step by step.}
\label{fig:follow}
\end{figure}

Since we stored all intermediate results, we can visualize the effect of each
intermediate step. We use \code{compare} and \code{cells} to follow the changes
in the data set as it is treated step by step. The \code{cells} function is
also part of \code{validate}. It measures the number of changes by decomposing
the number of cells in a data set into those that are available or not, and
separating those again into which are still available, have been removed, or
are still missing when compared to a previous version of the data set.
(See Figure~\ref{fig:follow}).
\begin{Schunk}
\begin{Sinput}
R> plot(compare(rules, raw= dat_raw, modified = dat_mod
+   , errors_located = dat_el, imputed = dat_imp
+   , adjusted = dat_adj, how="sequential"))
\end{Sinput}
\end{Schunk}
\begin{Schunk}
\begin{Sinput}
R> plot(cells(raw= dat_raw, modified = dat_mod
+   , errors_located = dat_el, imputed = dat_imp
+   , adjusted = dat_adj, compare="sequential"))
\end{Sinput}
\end{Schunk}
These \code{plot} methods can provide some quick first insight into the effect
that each step has on the dataset, thus providing potentially useful
information for improving the process. For example, in the left panel we see
that across the procedure, the number of violations (thick red line) first
decreases to zero after error localization, then errors get introduced by
imputing without regard of the restrictions, and finally all constraints are
satisfied by adjusting the imputed values. In the right panel we see that the
number of missings (thick yellow line) increases at error localization, reduces
to zero after imputation (everything can be imputed) and stays zero afterwards.
Such displays can help when assessing the performance of a data cleaning
process, for example by switching imputation methods or altering the modifying
rules. For cases where there is no natural ordering in a sequence of data sets,
the \code{barplot} method has also been overloaded for objects of classes

Summarizing, we see that \pkg{validate} plays an integral role in the
data cleaning procedure demonstrated here. Not only is it used to define
validation rules and to monitor data quality as a function of 
processing step, it also provides control parameters for two crucial
data cleaning steps: error localization and adjusing numerical values.

\section{Summary and outlook}
\label{sect:conclusion} 
The \pkg{validate} package demonstrated in this paper serves as a data
validation infrastructure in several aspects.  First, it is grounded in a
strict definition of the concept of data validation functions. This definition
is supported in the form of an \proglang{R}-embedded domain specific language
that allows users to express restrictions on data (validation rules) with great
flexibility. Second, the package treats validation rules as first class
citizens, implementing full CRUD (create, read, update, delete) functionality
on validation rule sets along with rudimentary rule inspection functionality.
Third, a flexible file-based interface makes rule sets and their metadata
elements maintainable as source code, complete with recursive file inclusion
functionality. The ability to reading rules and their metadata from a data
frame also allows storage and retrieval from relational databases.  Fourth, and
finally, the package is implemented using a fairly simple, extendable
object-oriented framework. This allows extensions in the direction of different
data sources and different rule set implementations.

Future work on this package will likely include improved support for subsetting
validation rule sets based on metadata elements, support for user-defined
extensions of the validation DSL, more helper functions for complex validation
tasks, and improved support for summarization and reporting of data validation
results.

\bibliography{jss3483}

\begin{thebibliography}{31}
\newcommand{\enquote}[1]{``#1''}
\providecommand{\natexlab}[1]{#1}
\providecommand{\url}[1]{\texttt{#1}}
\providecommand{\urlprefix}{URL }
\expandafter\ifx\csname urlstyle\endcsname\relax
  \providecommand{\doi}[1]{doi:\discretionary{}{}{}#1}\else
  \providecommand{\doi}{doi:\discretionary{}{}{}\begingroup
  \urlstyle{rm}\Url}\fi
\providecommand{\eprint}[2][]{\url{#2}}

\bibitem[{Armstrong(1974)}]{armstrong1974dependency}
Armstrong W (1974).
\newblock \enquote{Dependency Structures of Data Base Relationships.}
\newblock In \emph{IFIP Congress}, pp. 580--583. IFIP.

\bibitem[{Bache and Wickham(2014)}]{bache2014magrittr}
Bache SM, Wickham H (2014).
\newblock \emph{magrittr: A Forward-Pipe Operator for \proglang{R}}.
\newblock \proglang{R} package version 1.5,
  \urlprefix\url{https://CRAN.R-project.org/package=magrittr}.

\bibitem[{Beeri \emph{et~al.}(1984)Beeri, Dowd, Fagin, and
  Statman}]{beeri1984structure}
Beeri C, Dowd M, Fagin R, Statman R (1984).
\newblock \enquote{On the Structure of Armstrong Relations for Functional
  Dependencies.}
\newblock \emph{Journal of the ACM (JACM)}, \textbf{31}(1), 30--46.

\bibitem[{Bohannon \emph{et~al.}(2007)Bohannon, Fan, Geerts, Jia, and
  Kementsietsidis}]{bohannon2007conditional}
Bohannon P, Fan W, Geerts F, Jia X, Kementsietsidis A (2007).
\newblock \enquote{Conditional Functional Dependencies for Data Cleaning.}
\newblock In \emph{Data Engineering, 2007. ICDE 2007. IEEE 23rd International
  Conference on}, pp. 746--755. IEEE.

\bibitem[{Breiman(1984)}]{breiman1984cart}
Breiman L (1984).
\newblock \emph{Classification and Regression Trees}.
\newblock Taylor \& {F}rancis {G}roup, New {Y}ork.

\bibitem[{Chambers(2016)}]{chambers2016extending}
Chambers JM (2016).
\newblock \emph{Extending \proglang{R}}.
\newblock The {R} {S}eries. Chapman \& {H}all.

\bibitem[{{de Jonge} and {van der Loo}(2018)}]{jonge2015editrules}
{de Jonge} E, {van der Loo} M (2018).
\newblock \emph{editrules: Parsing, Applying, and Manipulating Data Cleaning
  Rules}.
\newblock \proglang{R} package version 2.9.3,
  \urlprefix\url{https://CRAN.R-project.org/package=editrules}.

\bibitem[{de~Jonge and {van der Loo}(2019)}]{jonge2018validatetools}
de~Jonge E, {van der Loo} M (2019).
\newblock \emph{validatetools: Checking and Simplifying Validation Rule Sets}.
\newblock \proglang{R} package version 0.4.6,
  \urlprefix\url{https://CRAN.R-project.org/package=validatetools}.

\bibitem[{de~Waal(2017)}]{waal2017imputation}
de~Waal T (2017).
\newblock \enquote{Imputation Methods Satisfying Constraints.}
\newblock In \emph{{Work Session on Statistical Data Editing}}, {Working Paper
  5}. {United Nations Economic Commission for Europe}.

\bibitem[{de~Waal \emph{et~al.}(2011)de~Waal, Pannekoek, and
  Scholtus}]{waal2011handbook}
de~Waal T, Pannekoek J, Scholtus S (2011).
\newblock \emph{Handbook of Statistical Data Editing and Imputation}, volume
  563.
\newblock John Wiley \& Sons.

\bibitem[{{Di Zio} \emph{et~al.}(2015){Di Zio}, Fursova, Gelsema, Giessing,
  Guarnera, Ptrauskiene, von Kalben, Scanu, ten Bosch, van~der Loo, and
  Walsdorfe}]{zio2015methodology}
{Di Zio} M, Fursova N, Gelsema T, Giessing S, Guarnera U, Ptrauskiene J, von
  Kalben LQ, Scanu M, ten Bosch K, van~der Loo M, Walsdorfe K (2015).
\newblock \enquote{Methodology for Data Validation.}
\newblock \emph{Technical Report deliverable No.}, ESSNet on validation.
\newblock
  \urlprefix\url{https://ec.europa.eu/eurostat/cros/content/methodology-data-validation-handbook-final_en}.

\bibitem[{Fellegi and Holt(1976)}]{fellegi1976systematic}
Fellegi IP, Holt D (1976).
\newblock \enquote{A Systematic Approach to Automatic Edit and Imputation.}
\newblock \emph{Journal of the American Statistical Association},
  \textbf{71}(353), 17--35.

\bibitem[{Fischetti(2019)}]{fischetti2018assertr}
Fischetti T (2019).
\newblock \emph{assertr: Assertive Programming for \proglang{R} Analysis
  Pipelines}.
\newblock \proglang{R} package version 2.6,
  \urlprefix\url{https://CRAN.R-project.org/package=assertr}.

\bibitem[{Fowler(2010)}]{fowler2010domain}
Fowler M (2010).
\newblock \emph{Domain-Specific Languages}.
\newblock Pearson Education.

\bibitem[{Gibbons(2013)}]{gibbons2013functional}
Gibbons J (2013).
\newblock \enquote{Functional Programming for Domain-Specific Languages.}
\newblock In V~Zs\'ok, Z~Horv\'ath, L~Csat\'o (eds.), \emph{Central European
  Functional Programming - Summer School on Domain-Specific Languages}, volume
  8606 of \emph{LNCS}, pp. 1--28. Springer-Verlag.
\newblock \doi{10.1007/978-3-319-15940-9_1}.
\newblock
  \urlprefix\url{http://link.springer.com/chapter/10.1007/978-3-319-15940-9_1}.

\bibitem[{Iannone(2018)}]{iannone2018pointblank}
Iannone R (2018).
\newblock \emph{pointblank: Validation of Local and Remote Data Tables}.
\newblock \proglang{R} package version 0.2.0,
  \urlprefix\url{https://CRAN.R-project.org/package=pointblank}.

\bibitem[{Petersen and Ekstr{\o}m(2019)}]{petersen2018datamaid}
Petersen AH, Ekstr{\o}m CT (2019).
\newblock \enquote{dataMaid: Your Assistant for Documenting Supervised Data
  Quality Screening in \proglang{R}.}
\newblock \emph{Journal of {S}tatistical {S}oftware}, \textbf{90}, 1--38.

\bibitem[{{\proglang{R} Core Team}(2019)}]{rcore}
{\proglang{R} Core Team} (2019).
\newblock \emph{\proglang{R}: A Language and Environment for Statistical
  Computing}.
\newblock \proglang{R} Foundation for Statistical Computing, Vienna, Austria.
\newblock \urlprefix\url{https://www.R-project.org/}.

\bibitem[{Rumbaugh \emph{et~al.}(2004)Rumbaugh, Jacobson, and
  Booch}]{rumbaugh2004UML}
Rumbaugh J, Jacobson I, Booch G (2004).
\newblock \emph{Unified Modeling Language Reference Manual, The (2Nd Edition)}.
\newblock Pearson Higher Education.
\newblock ISBN 0321245628.

\bibitem[{{Statistics Netherlands}(2018)}]{cbs2018blaise}
{Statistics Netherlands} (2018).
\newblock \emph{{BLAISE}}.
\newblock Version 5.2, \urlprefix\url{https://blaise.com}.

\bibitem[{Tempelman(2007)}]{tempelman2007imputation}
Tempelman C (2007).
\newblock \emph{Imputation of Restricted Data}.
\newblock Ph.D. thesis, University of Groningen.

\bibitem[{{van der Loo}(2017)}]{loo2017simputation}
{van der Loo} M (2017).
\newblock \emph{simputation: Simple Imputation}.
\newblock \proglang{R} package version 0.2.3,
  \urlprefix\url{https://CRAN.R-project.org/package=simputation}.

\bibitem[{{van der Loo}(2019)}]{loo2018rspa}
{van der Loo} M (2019).
\newblock \emph{rspa: Adapt Numerical Records to Fit (in)Equality
  Restrictions}.
\newblock \proglang{R} package version 0.2.4,
  \urlprefix\url{https://CRAN.R-project.org/package=rspa}.

\bibitem[{{van der Loo} and {de Jonge}(2018)}]{loo2018dcmodify}
{van der Loo} M, {de Jonge} E (2018).
\newblock \emph{dcmodify: Modify Data Using Externally Defined Modification
  Rules}.
\newblock \proglang{R} package version 0.1.2,
  \urlprefix\url{https://CRAN.R-project.org/package=dcmodify}.

\bibitem[{van~der Loo and de~Jonge(2018)}]{loo2018statistical}
van~der Loo M, de~Jonge E (2018).
\newblock \emph{Statistical Data Cleaning with Applications in \proglang{R}}.
\newblock John Wiley \& Sons.

\bibitem[{{van der Loo} and {de Jonge}(2019)}]{loo2018validate}
{van der Loo} M, {de Jonge} E (2019).
\newblock \emph{validate: Data Validation Infrastructure}.
\newblock R package version 0.9.2,
  \urlprefix\url{https://CRAN.R-project.org/package=validate}.

\bibitem[{van~der Loo and de~Jonge(2020)}]{loo2019data}
van~der Loo M, de~Jonge E (2020).
\newblock \enquote{Data Validation.}
\newblock \emph{Wiley StatsRef}.
\newblock Accepted for publication.

\bibitem[{Vink(2015)}]{vink2015restrictive}
Vink G (2015).
\newblock \emph{Restrictive Imputation of Incomplete Survey Data}.
\newblock Ph.D. thesis, Utrecht University.
\newblock
  \url{https://dspace.library.uu.nl/bitstream/handle/1874/308699/vink.pdf}.

\bibitem[{{VTL task force} and {SDMX technical working
  group}(2018)}]{sdmx2018vtl}
{VTL task force}, {SDMX technical working group} (2018).
\newblock \emph{{Validation and Transformation Language}}.
\newblock Version 2.0, \urlprefix\url{https://sdmx.org}.

\bibitem[{yaml.org(2015)}]{yaml2015}
yamlorg (2015).
\newblock \enquote{{YAML} {A}int {M}arkup {L}anguage.}
\newblock {h}ttp://yaml.org/ (accessed 2015-08-13).

\bibitem[{Zhang and Pannekoek(2015)}]{zhang2015optimal}
Zhang LC, Pannekoek J (2015).
\newblock \enquote{Optimal Adjustments for Inconsistency in Imputed Data.}
\newblock \emph{Survey methodology}, \textbf{41}(1), 127--144.

\end{thebibliography}
\end{document}